\documentclass[fleqn,usenatbib]{mnras}

\usepackage{newtxtext,newtxmath}

\usepackage[T1]{fontenc}
\usepackage{ae,aecompl}

\usepackage[normalem]{ulem}
\usepackage{amsmath}
\usepackage{graphicx}
\usepackage{dcolumn}
\usepackage{bm}
\usepackage{epsfig}
\usepackage{latexsym}
\usepackage{color,colortbl} 
\usepackage[usenames,dvipsnames]{xcolor}
\usepackage{subfigure,float}

\usepackage{multirow}
\usepackage{textcomp}
\usepackage{array}
\usepackage[normalem]{ulem}
\usepackage{hyperref}
\usepackage{ulem}
\usepackage{verbatim}
\usepackage{url}
\hypersetup{
	colorlinks=true,
	linkcolor=red,
	citecolor=blue,
} 
\definecolor{LightCyan}{rgb}{0.88,1,1}

\usepackage{color,soul}

\newcommand{\be}{\begin{equation}}
\newcommand{\ee}{\end{equation}}
\newcommand{\bs}{\begin{split}} 
\newcommand{\bea}{\begin{eqnarray}}
\newcommand{\eea}{\end{eqnarray}}
	
\newcolumntype{M}[1]{>{\centering\arraybackslash}m{#1}}
	
\newcommand{\dtbf}{\Delta t_{\rm fit}}
\newcommand{\dtt}{\Delta t_{\rm true}}
\newcommand{\dt}{\Delta t} 
\newcommand{\dti}{\Delta t_i} 

\volume{511}
\pagerange{1210--1217}
\pubyear{2022}

\title{Out of One, Many: Distinguishing Time Delays from Lensed Supernovae} 

\author[M. Denissenya et al.]{
	Mikhail Denissenya,$^1$\thanks{Email: mikhail.denissenya@nu.edu.kz}
	Satadru Bag$^2$, Alex G.~Kim$^3$, Eric V.~Linder$^{1,3,4}$,	Arman Shafieloo$^{2,5}$ 
\\
\\
$^1$ Energetic Cosmos Laboratory, Nazarbayev University, Nur-Sultan 010000, Kazakhstan\\ 
$^2$ Korea Astronomy and Space Science Institute, Daejeon 34055, Korea\\ 
$^3$ Lawrence Berkeley National Laboratory, Berkeley, CA 94720, USA\\ 
$^4$ Berkeley Center for Cosmological Physics, University of California, Berkeley, CA 94720, USA\\
$^5$ Department of Astronomy and Space Science, University of Science and Technology, Daejeon 34113, Korea
} 

\begin{document}
\label{firstpage}

\maketitle

\begin{abstract}
Gravitationally lensed Type Ia supernovae are an emerging probe 
with great potential for constraining dark energy, spatial curvature, 
and the Hubble constant. The multiple images and their time delayed 
and magnified fluxes may be unresolved, however, blended into a 
single lightcurve. We demonstrate methods without a fixed source 
template matching for extracting the 
individual images, determining whether there are one (no lensing) or 
two or four (lensed) images, and measuring the time delays between 
them that are valuable cosmological probes. We find 100\%  success for determining the number of images for time delays 
greater than $\sim10$ days. 

\end{abstract}


%%%%%%%%%%%%%%%%%%%%%%%%%%%%%%%%%%%%%%%%%%%%%%%%%%

%%%%%%%%%%%%%%%%% BODY OF PAPER %%%%%%%%%%%%%%%%%%

\section{Introduction}
Time delays in strong gravitational lens systems 
are one of the few dimensionful quantities that 
can be measured cosmologically. This enables an absolute 
distance measurement, rather than a ratio of distances, and 
so determines an absolute cosmic length scale, e.g.\ the 
Hubble constant \citep{refsdal}. The time delay distance formed 
from the time delay involves combinations of distances from 
observer to source, observer to lens, and lens to source; 
this is a valuable probe of both the cosmic expansion 
behavior and hence dark energy, and spatial curvature 
\citep{linder04,linder11,wong,millon,shajib,liao,treumar}.

Each strongly lensed source is split so as to appear as 
multiple images, with generally the central lensed image 
unobservable due to lens galaxy obscuration, leaving two or four images. These images will 
be magnified in flux, and delayed in time, relative to the 
source flux and each other. 
Type Ia supernovae (hereafter just SN) have particular advantages as gravitationally 
lensed sources, as they are time variable (so time delays are 
detectable), their intrinsic time variation is fairly well 
known, the observations needed to measure time delays 
span a month or two rather than years as for lensed quasars, 
and the SN distances are standardizable, simplifying the lens 
system modeling.

If the multiple images can be resolved, i.e.\ their angular 
separations are greater than the observing resolution, 
then each image can be measured separately. 
Some particular recent work on such lensed supernova cosmology appears in,  e.g., 
\citep{2108.02789,2106.08935,2010.12399}. 
However, for worse image resolution, 
smaller lensing masses, or less favorable geometry of the 
lens and source positions, the images may overlap and blend 
together -- be unresolved. In  this case only a single combined 
lightcurve (flux vs time) can be measured, defining a system. We only use this observed lightcurve data. 

We focus here on such unresolved lensed SN, investigating whether 
we can determine from a combined lightcurve its constituent 
elements: is it lensed or unlensed, how many images, and what 
are their time delays (and magnifications, to a lesser extent). For unresolved lensed supernovae this work covers the key range of 6--14 days motivated by the expected distribution of time delays generally peaked around 10 days as in, e.g., \citep{goldstein}. 
This extends Paper 1 \citep{paper1} in this series by going 
beyond two image systems to allow four images and a more 
robust exploration of multiplicity, and using a complementary  
fitting method, also without a fixed source template. 

In Section~\ref{sec:method} we describe the new fitting 
method and crosscheck it against both simulations and the previous 
method for two images. We analyze it for four image systems in 
Section~\ref{sec:image4} and quantify its precision, accuracy, 
and range of robustness against simulations. Section~\ref{sec:multi} 
investigates determination of multiplicity, i.e.\ distinguishing 
unlensed (one image) systems from two image from four image systems, 
presenting the ``confusion matrix'' between them. We summarize 
the current state and future work in fitting unresolved lensed 
SN in Section~\ref{sec:concl}.

%%%%%%%%%%%%%%%%%%%%%%%%%%% 
\section{Freeform Fitting} \label{sec:method} 

Strong lensing creates multiple images, each delayed in time and 
magnified (greater or less) in flux. If the images can not be 
resolved, then the signal is blended, a superposition of all 
images. The resulting blended lightcurve defines a system. For an intrinsic, unlensed source flux $U(t)$, the 
measured, blended lightcurve is 
\be 
F_j(t)=\sum_{i \in \rm images} \mu_i\,U_j(t-\Delta t_i)\,. 
\ee 
Here $j$ indexes different wavelength filter bands, $i$ the 
different images, and $\mu_i$ and $\Delta t_i$ give the 
magnification and time delay, respectively, of image $i$. 
We aim to fit for $\Delta t_i$ and $\mu_i$ (though we are mostly 
interested in $\Delta t_i$); note these quantities are one (important) step among many to use  strong lensing systems for cosmology. The freeform technique works strictly from the observed lightcurve data: supernova properties such as stretch, color, redshift do not enter further as the method is freeform, i.e.\ model agnostic. As in Paper 1, we do not include 
microlensing, which would turn the numbers $\mu_i$ into 
time and wavelength dependent functions $\mu_{i,j}(t)$. We 
address microlensing in future work, but here focus on a direct 
comparison to and extension of Paper 1. 

If one knew the source time evolution $U(t)$ perfectly, then 
fitting for $\dti$, $\mu_i$ is straightforward. However, SN 
do vary in their lightcurve intrinsically, and any template 
adopted will lead to errors, and possibly biases, in the 
extracted time delays. Furthermore, we wish to keep our method 
reasonably general so it might be applicable in the future to 
other lensed transients besides Type Ia supernovae. Therefore 
we did not adopt a fixed template in Paper 1, and do not here.

%%%%%%%%%%%%%%%% 
\subsection{Free Within Bounds}  \label{sec:methoddetail} 

In Paper 1, we took a base lightcurve form with an asymmetric 
rise and fall (specifically a lognormal) and multiplied it with a Crossing hyper-function (fourth order polynomial constructed from the first four Chebyshev 
polynomials) \citep{2011JCAP...08..017S,2012JCAP...05..024S,2012JCAP...08..002S,2014JCAP...01..043H}. This gives considerable freedom for the lightcurve 
shape. However, it does correlate the shape at one time with 
that at another time (recall the Chebyshev polynomials represent 
a linear tilt, parabolic curvature, etc.\ as they increase in order). 
Therefore we also explored an alternate form allowing freedom 
in the lightcurve shape, which we use in this paper. 

The technique here is of freedom at each time (SN phase), 
uncorrelated with other times. Complete freedom in the lightcurve 
shape would lead to a fully degenerate problem, so we impose 
bounds on the amplitude of the freedom. Concretely, we have 
\be 
U_j(t)=H_j(t) \left[ 1+h_j(t_k)\right]_{GS(t)}\,,
\label{eq:unlens}
\ee 
where $H_j(t)$ is the Hsiao \citep{hsiao} template for Type Ia 
supernovae in wavelength band $j$. We use 
Zwicky Transient Facility (ZTF; \citet{ztf}) $g$, $r$, $i$ bands. 
The fit parameters $h_j$ are hyperparameters embodying the 
freedom to change the lightcurve shape, but we bound their amplitude 
by $|h_j|<b$. We explored various values for the limiting fractional 
change $b$, and found that $b=0.1$, i.e.\ a 10\% change was ample to cover the variation of SN intrinsic lightcurves 
used for cosmology, giving the best combination of freedom and accuracy.  

The hyperparameters $h_j$ are independently chosen at nodes $t_k$ 
throughout the SN phase. The lightcurve 
in between nodes, i.e.\ the full $U_j(t)$, is formed by 
multiplying a Gaussian kernel smoothing in time of the $1+h_j(t_k)$ factor 
(denoted by the subscript $GS(t)$; the smoothing length is itself a hyperparameter, lying in [3,8] days) 
by a linear spline of $H_j$ (the Hsiao template is 
defined at discrete epochs). We found this works better 
and is more computationally efficient than Gaussian 
smoothing (or splining) both factors. 
The number of nodes used is typically 50-70, 
depending on the observation range. Too small a 
number of nodes will not give sufficient freedom 
for variations, while too large a number of nodes 
will oversample the data as well as increase the 
computational cost. We experimented with both 
adaptive and equal time spacing, and found equal 
spacing to work well, avoiding the potential for 
adaptive spacing to become too sparse in certain 
epochs and miss useful features (e.g.\ troughs as 
well as peaks inform the time delay estimation). 

As in Paper 1, we simulate data using our \textsc{LCsimulator} 
code based on the \textsc{sncosmo} \citep{sncosmo} \textsc{python} package, 
with lightcurves generated by applying noise and 
observing characteristics as in \citep{goldstein} to 
Hsiao spectra (see Appendix~\ref{sec:apxsalt} for 
tests with alternative lightcurves). The modeled lightcurves in $g$, $r$, $i$ ZTF bands include observations with $\sim2-4$ day cadences with the flux uncertainties set to 5\% of the peak value.
We normalize the input data to unit intervals in 
flux and time for consistent treatment of numerical precision. 
For a system modeled with $N_\mathrm{image}$ images, then the $N_\mathrm{image}-1$
relative time delays and $N_\mathrm{image}-1$ relative magnifications are 
derived from $N_\mathrm{image}$ absolute time positions and $N_\mathrm{image}$ magnifications. 

We infer the time delays and relative magnifications using the No-U-Turn Hamiltonian Monte Carlo (HMC) sampler within \textsc{CmdStan}, the command-line interface to the Stan statistical modeling language \citep{stan}. Our typical HMC run includes 8 chains, each with 1400 iterations and 250 warm up steps, with a system taking $\sim0.4-4$ hours per chain, depending on the number of parameters (images) fit.

%%%%%%%%%%%%%%%%%%%%%%%% 
\subsection{Comparison with Crossing Method} \label{sec:crosscompare} 

Advantages exist for each one of the fitting methods we apply 
to lensed lightcurves. The method of Paper 1, which we call the 
crossing method, allows the amplitude of deviation from the 
base form to vary widely, but places some constraints on the shape 
deviations. The freeform method here allows the shape of the 
lightcurve to vary widely but bounds the amplitudes of deviation. 
Allowing both shape and amplitude to vary without constraint is 
untenable as strong degeneracies are introduced such that, for 
example, an unlensed SN could be made to look like a blend of two 
images. We believe both the freeform and crossing method are useful, 
and provide valuable crosschecks.

We begin the comparison between the two by considering the two-image systems used 
in Paper 1 for blind testing. By fitting them with the freeform 
technique we can directly compare the results. 

\begin{figure*}
	\centering
	
	\includegraphics[width=0.46\textwidth]{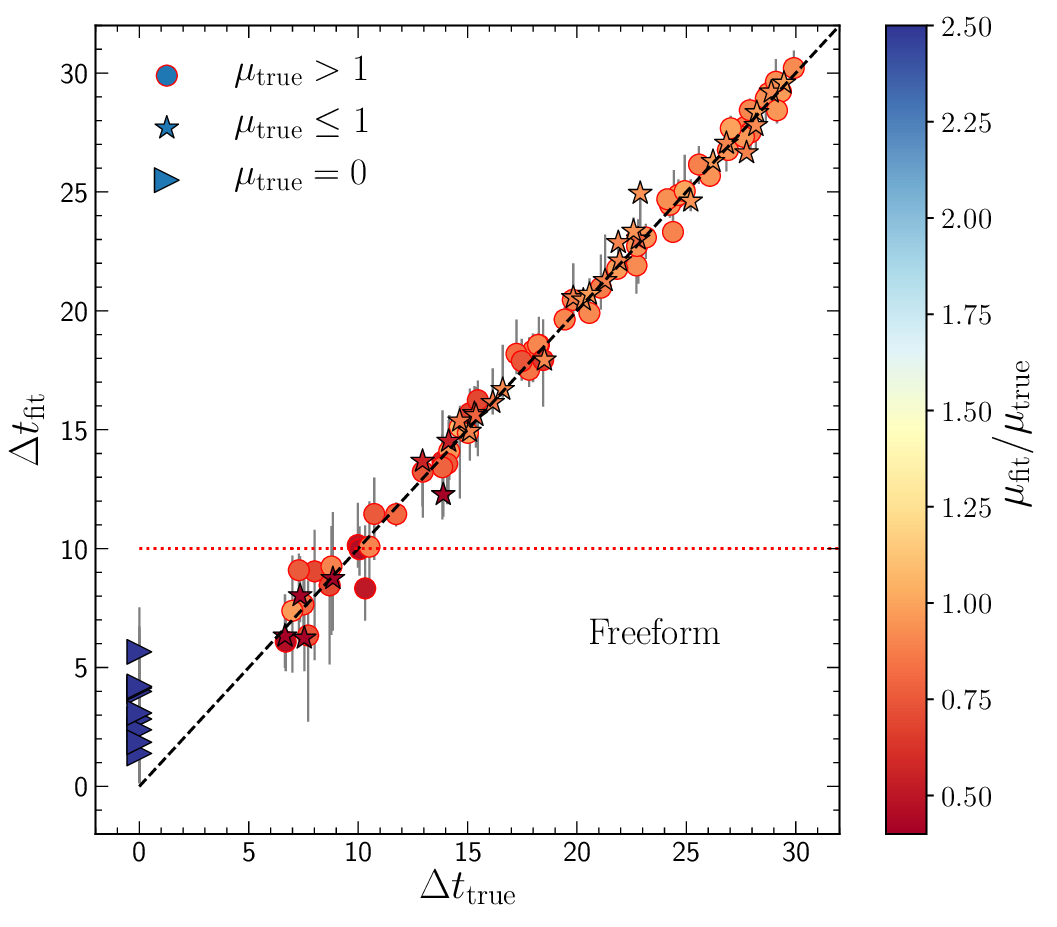}
	\includegraphics[width=0.46\textwidth]{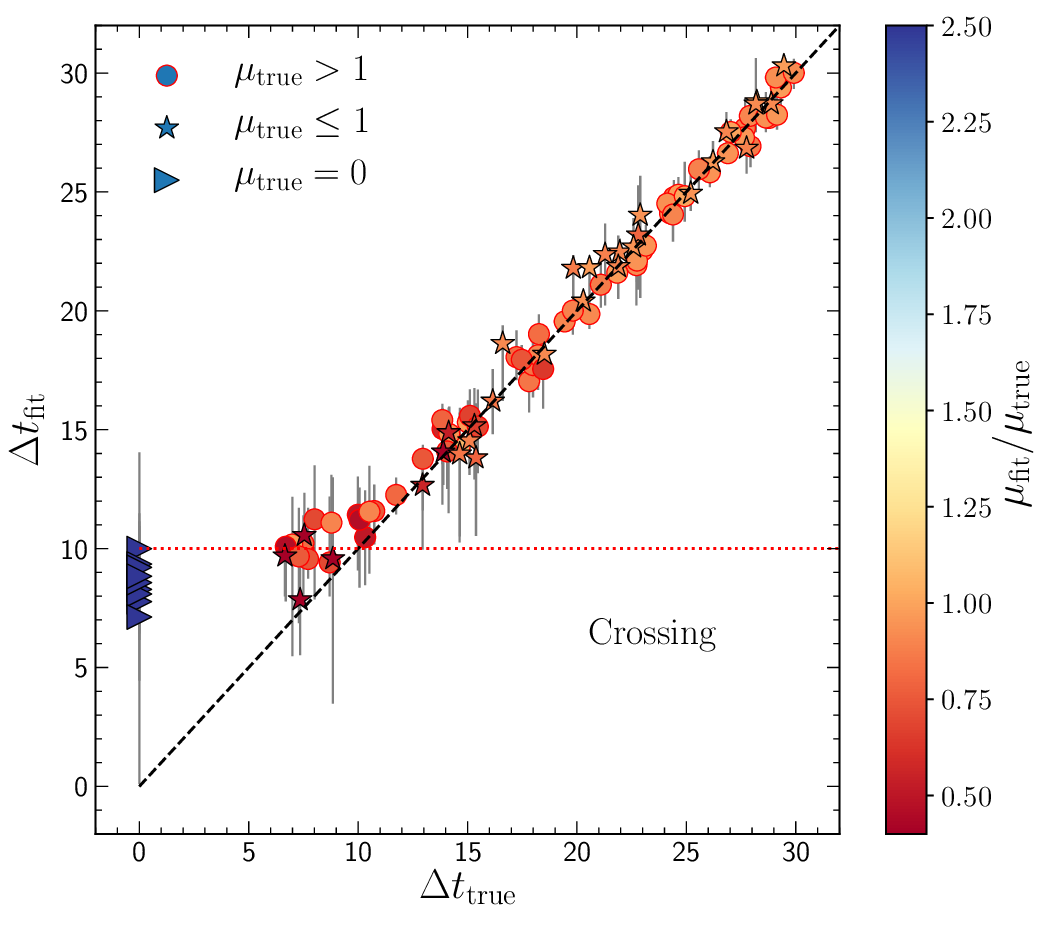}\\
	%\hspace{-40pt}
	\includegraphics[width=0.41\textwidth]{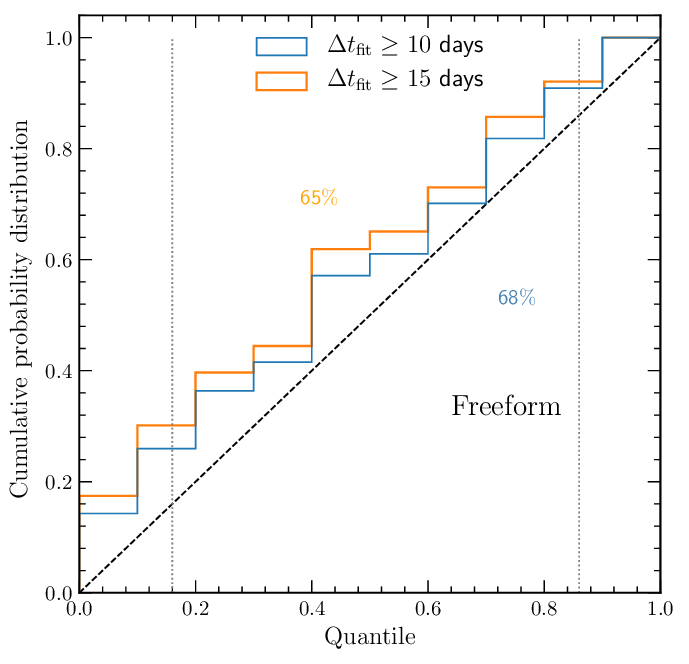}
	%\hspace{20pt}
	\includegraphics[width=0.41\textwidth]{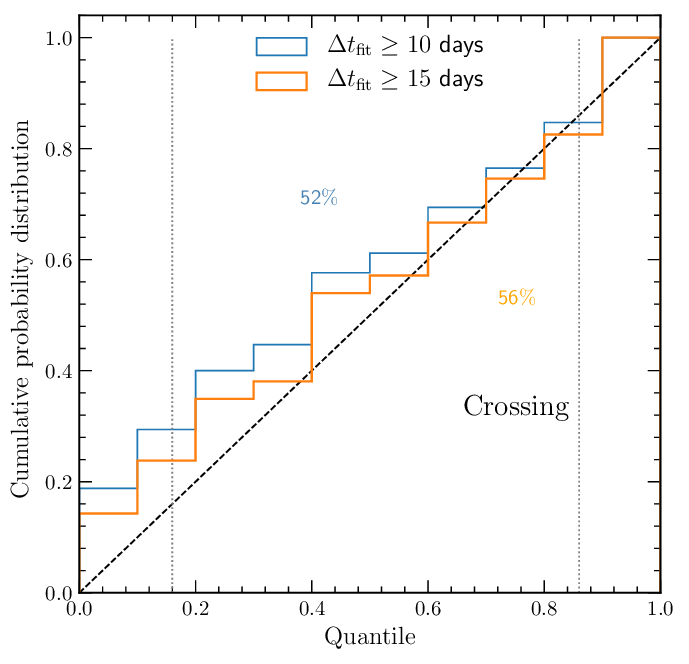}\\
	%\hspace{-40pt}
	\includegraphics[width=0.41\textwidth]{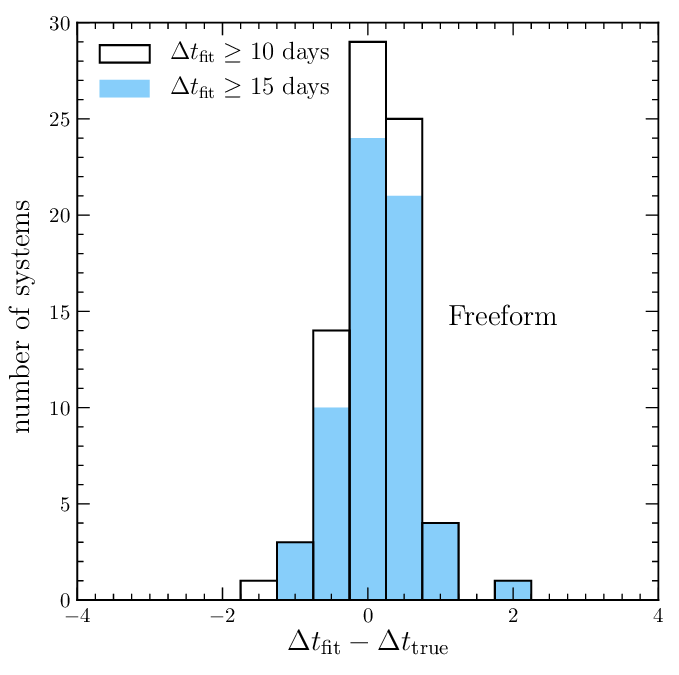}
%	\hspace{20pt}
	\includegraphics[width=0.41\textwidth]{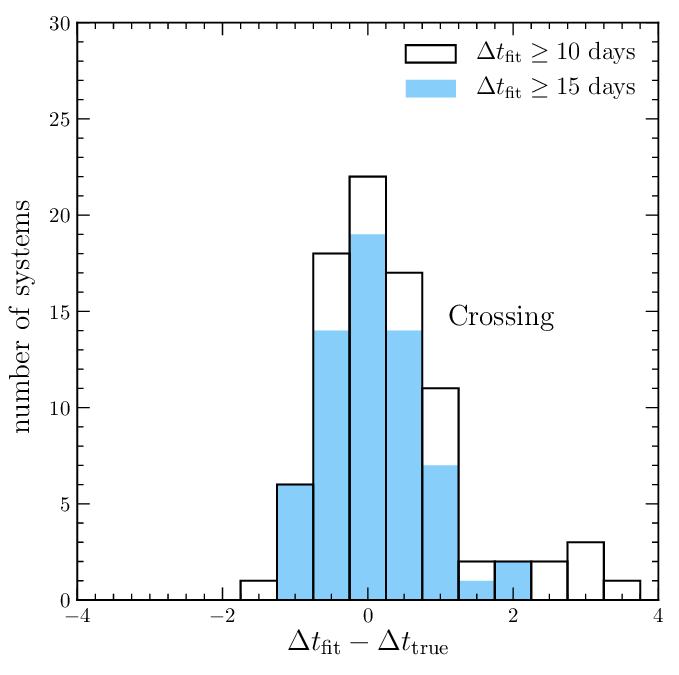}
	
	\caption{Blind tests using the freeform method (left 
		panels) and crossing technique (right panels) on a 
		set of $100$ systems, with 9 unlensed and 91 two image 
		systems with time delays are distributed over $6-30$ days. 
		[Top panels] Time delay fits lie closely along 
		the diagonal $\dtbf=\dtt$. For $\dtbf<10$ days, the 
		crossing technique of Paper 1 tends to level out, while 
		the freeform method continues to follow the diagonal. 
		Unlensed systems ($\dtt=0$) are also clearly distinct 
		from $\dtbf\ge10$ fits with the freeform method. Error bars show the 95\% confidence intervals.
		[Middle panels] The cumulative distribution functions 
		are shown for the systems with $\dtbf\ge10$ (blue) 
		and $\dtbf\ge15$ (orange) days. The central 68\% 
		of fit probability encompasses 65-68\% probability 
		in the freeform case, showing excellent accuracy, an 
		improvement over the 52-56\% of the crossing technique. 
		[Bottom panels] Histograms of the time delay fits vs 
		truth. Both show good peakiness around the truth. 
	} 
	\label{fig:ablindtest}
\end{figure*}

Figure~\ref{fig:ablindtest} shows the results for 
this paper's freeform technique (left) and Paper 1's 
crossing technique (right) for those 100 systems. 
The top panels show $\dtbf$ vs $\dtt$, the 
middle panels show 
the cumulative distribution function plots, and the 
lower panels the $\dtbf-\dtt$ 
histograms. 

%%%%%%%%%%%%%%%%%%%%%%%%% 

For $\dtbf$ vs $\dtt$, both techniques do quite well for 
$\dtbf\ge10$ days, tightly following the diagonal, i.e.\ truth. The 
freeform method continues to follow the diagonal below 
10 days, down to about 6 days (this was the limit of the Paper 1 testing; in 
Appendix~\ref{sec:aless10} here we see it actually works well to 2 days), while the crossing technique 
had leveled off with $\dtbf=10$  (all $\Delta t$ are given in units of days). Also, for the unlensed 
systems ($\dtt=0$), the freeform method seems to control the 
fit somewhat better, keeping false positives well below 
$\dtbf=10$, unlike the crossing technique. This makes the 
freeform method appear promising for application to 
the four image fits of 
the next section. 

For the cumulative distribution functions (CDF), the key 
characteristic is whether the distribution has slope one, 
i.e.\ the fit distribution follows the true distribution, 
in the central part of the diagram, 
indicating reliable fit uncertainties. We indicate the central 
68\% probability distribution, between the 0.16 and 0.84 
quantiles, by the vertical dotted lines. In that central  
68\% probability, the freeform method fits occur 65-68\% 
of the time (depending on the cutoff in $\dtbf$), demonstrating its accuracy, while 
the crossing technique had 52-56\%. Note that the vertical offset 
from the diagonal is due to a sample selection bias due to 
the imposed cutoff in $\dtbf$, so that quantile zero is shifted to   
CDF $>0$; but again, the central part of the diagram is the most important. 

The histograms of $\dtbf-\dtt$ in Fig.~\ref{fig:ablindtest} 
for $\dtbf\ge10$ days 
show that the freeform approach accurately recovers the true time delays, with better than 
1.25 day precision in $\sim$97\% of the cases,  
with accuracy $\lesssim0.1$ day. 
The freeform nature of the approach allows greater flexibility in the 
lightcurve shape for robustness and 
eliminates the bias seen for the crossing method when 
$10\leq \dtbf<15$. 

Thus the freeform technique has shown good accuracy and 
precision on two image systems, motivating its use for 
four image systems as well. We note that 
both approaches have advantages in particular areas. 
The crossing technique from Paper 1 starts from a form 
that is merely a rise and fall of flux, i.e.\ a fairly 
generic transient, without assuming it comes from a SN Ia,  
while the freeform technique does use deviations around 
a Hsiao SN Ia lightcurve. We might expect the crossing  
technique to be useful for unclassified transients, and then 
the freeform method to be more accurate for those suspected 
of being SN Ia. In the remainder of the paper we assume we 
are dealing with potentially lensed (one, two, or four 
images) SN Ia and employ the freeform method.

%%%%%%%%%%%%%%%%%%%%%%%%%%% 
\section{Time Delays of Four Images} \label{sec:image4} 

Having established that the freeform method works well on two 
image systems, and distinguishes them from unlensed SN, we 
now proceed further and consider four image systems (recall that 
strong lensing generally produces two or four visible images, 
not an odd number). 

For four image systems there are four (unobservable) times 
$t_1$, $t_2$, $t_3$, $t_4$ associated with the four images, 
and three observable, independent 
relative time delays, $\dt_{AB}$, $\dt_{BC}$, $\dt_{CD}$. We 
use the numerical 1-4 and alphabetical $A-D$ 
notation to emphasize the difference between 
unobservables and observables. Furthermore, there are four 
magnifications relative to the unlensed source flux, so there 
are a large number of permutations for the test system 
parameters. 
We therefore carry out two types of studies, one with 
systematic time delays and one with random time delays within 
a range. 

First we systematically study the accuracy of four image 
fits  by considering equal true delays:  
$\dt_{AB}=\dt_{BC}=\dt_{CD}\equiv \Delta T$ for $\Delta T=4$, 6, 8, 
10, 12 days. We still fit for each time delay 
as independent free parameters. These systems are simulated with  
a flux noise level of 5\% of peak flux and a 
variety of image magnifications. 

\begin{figure*}
	\centering
	\includegraphics[width=0.32\textwidth]{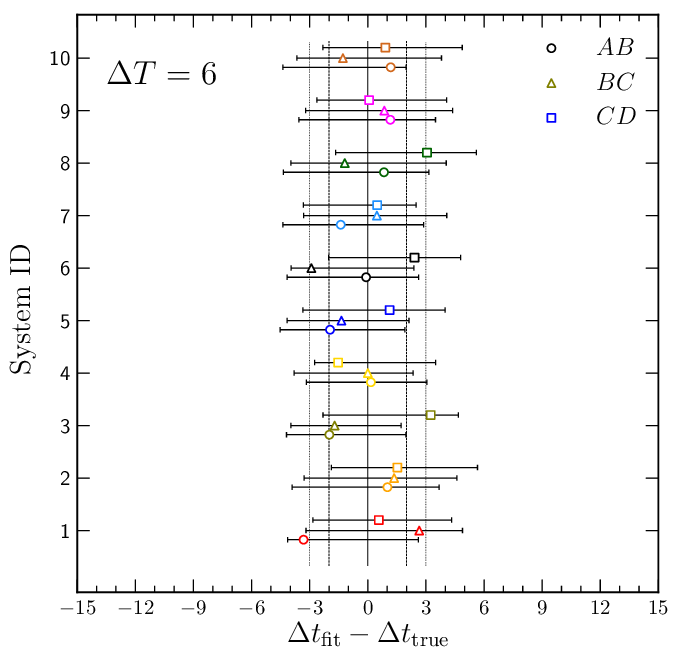}
	\includegraphics[width=0.32\textwidth]{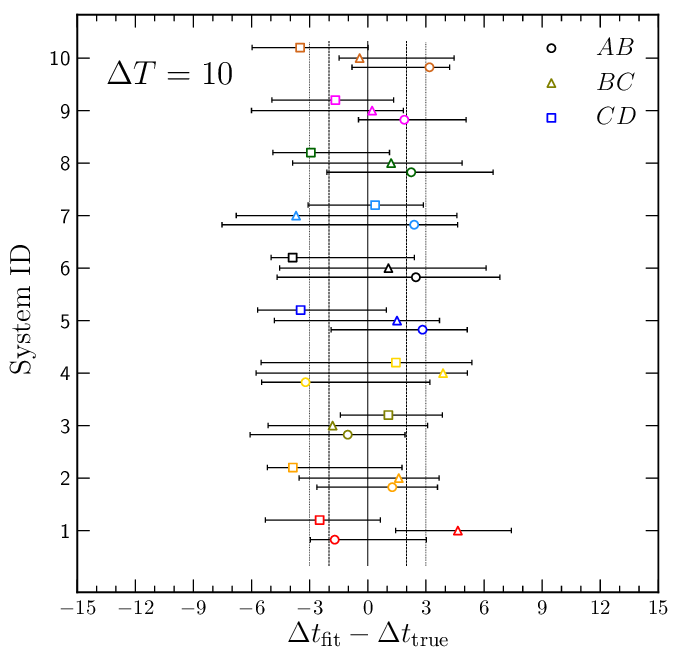}
	\includegraphics[width=0.32\textwidth]{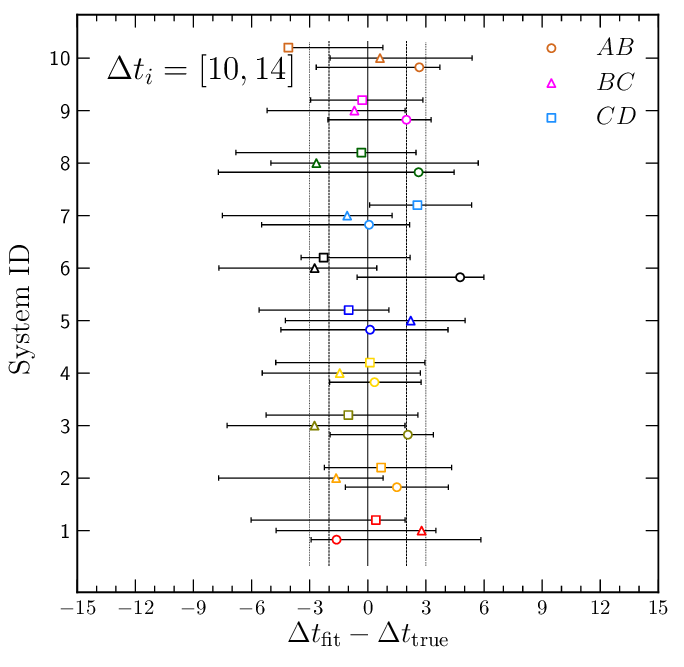}
	\caption{The time delay deviations from the truth in four-image systems for the equal time delay differences $\Delta T=6$ (left panel), $\Delta T=10$ (middle panel), and randomly chosen from the $\dti=[10,14]$ interval 
		(right panel). The thin dashed and dotted vertical lines are used to indicate 2 and 3 day deviations respectively. Error bars show the  
		95\% confidence intervals. 
	} 
	\label{fig:sp4im}
\end{figure*}

Figure~\ref{fig:sp4im} shows the 
fits, relative to truth, for the three 
relative time delays $\dt_{AB}$, $\dt_{BC}$, $\dt_{CD}$ 
for each of ten four-image systems (we choose ten systems randomly, enough to be statistically informative while not swamping the figure with 100+ data bars). We exhibit the 
cases for $\Delta T=6$ (left panel) and $\Delta T=10$ (middle panel). 
While the fits have more variation than those for two images, 
nearly all contain the true value within the 95\% 
uncertainties. 
Even for $\Delta T=6$ (well below what we will use), 
there is clear recognition that it is lensed at some 
nonzero time delay.

%%%%%%%%%%%%%%%%%%%%%%%%%%%%%%%%%%%%% 

While we can fit four images well, plus distinguish them 
from the unlensed (zero time delay) case, we also want to 
see if we can distinguish them from the two image, lensed 
case. We therefore take the same systems 
and fit them as two image systems as well, i.e.\ with a 
single relative time delay and a relative magnification. 

\begin{figure}
	\centering
	\includegraphics[width=\columnwidth]{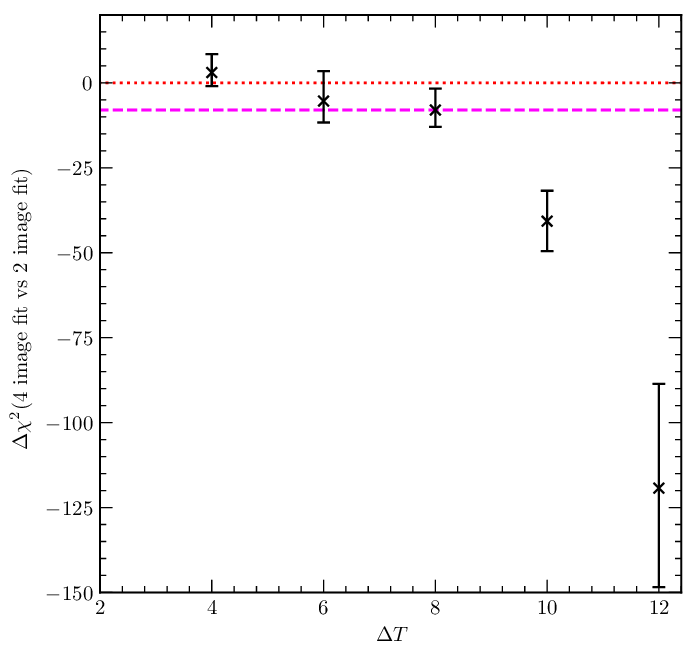} 
	\caption{Distinction between systems with four images and with two images is statistically highly significant for 
		time delay spacing of greater than eight days, generated for four image systems. Error bars give the mean $\Delta\chi^2$ difference and 68\% confidence region at each $\Delta T$ spacing. 
		Fits with $\Delta\chi^2<0$ (dotted red line) correctly prefer a four image fit, while those with $\Delta\chi^2<-8$ (dashed magenta line) have Akaike Information Criterion AIC $<0$, accounting for the four extra parameters of a four vs two image fit, and hence are fully robust. 
	}
	\label{fig:chi4vs2T}
\end{figure}

Figure~\ref{fig:chi4vs2T} evaluates the relative $\chi^2$ 
of the fit with four images relative to the fit with two 
images, as a function of true (four images) time delay 
spacing $\Delta T$. If $\Delta\chi^2\equiv\chi^2_{\rm min,4im}-\chi^2_{\rm min,2im}<0$, this means the four image fit 
is preferred. 
However, as the four image fit has four more 
parameters (two more time delays and two more relative 
magnifications), one might also look at the Akaike Information 
Criterion AIC $=\Delta\chi^2-2\Delta N_{\rm dof}$. Then 
four images is robustly preferred when $\Delta\chi^2<-8$. 
These two criteria are shown as the dotted red and dashed 
magenta horizontal lines, respectively. We see that for 
$\Delta T>8$ days we can have significant confidence that we 
can distinguish four image from two image lensed systems.

%%%%%%%%%%%%%%%%% 

For the second study, we focus on time delays randomly 
distributed in the range $\dti=[10,14]$, 
where the subscript runs over $AB$, $BC$, $CD$. 
While we have 
seen that we can fit well smaller time delays for four 
image systems, and for two image systems, we will not have 
as robust confidence that we can tell the two cases apart. 
Therefore we concentrate on this range, which should be the 
most challenging of the fits for which we have confidence not 
only in the fits themselves but the number of images as well 
(see Section~\ref{sec:multi} for further investigation). 
We also randomly select the magnifications $\mu_i$. 

We set flat priors on $t_1$ and $t_4$. 
The lower bound on 
$t_1$ is given by when the flux in $g$ band reaches 15\% 
of flux peak value, and the upper bound is determined from 
the flux peak position. Similarly, the upper bound on $t_4$ 
is when the flux reaches the 15\% of the peak value on 
the tail of the observation interval. The bounds on $t_2$ and  
$t_3$, and the lower bound on $t_4$, are determined 
dynamically in the fit, subject to the inequalities 
\be
t_2>\tau t_1,\quad t_3>\tau t_2,\quad t_4>\tau t_3, 
\ee
where the factor $\tau>1$ is used to preserve some 
separation between the images. That is, we don't want 
images lying right on top of each other and hence 
being degenerate with a lesser number of images. 
We find that $\tau=1.1$ works well.  

For the magnification  
we impose $\mu_i>0.25$, since recall that $\mu_i=0$ means 
there is no image (so a four image system with two $\mu_i=0$ is 
really a two image system) and there is a perfect degeneracy for 
all values of that $\dti$, while for $\mu_i$ near zero such 
a low amplitude bump is degenerate with a local shape 
perturbation $h(t)$. 
Thus the condition $\mu_i>0.25$ is employed to avoid these issues 
and enable robust fit convergence. The upper bound is $\mu_i<4$. 

\begin{figure*}
	\centering
	\includegraphics[width=0.47\textwidth]{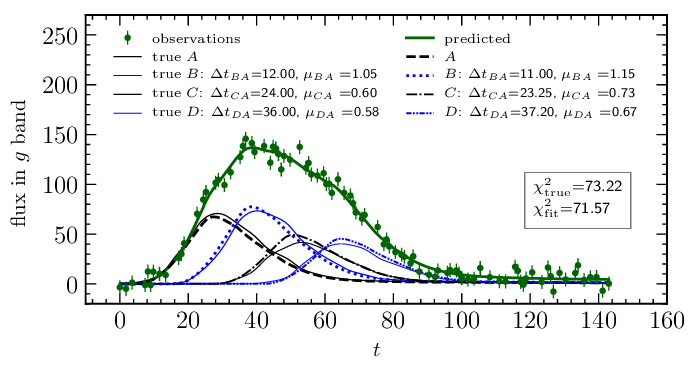}
	\includegraphics[width=0.47\textwidth]{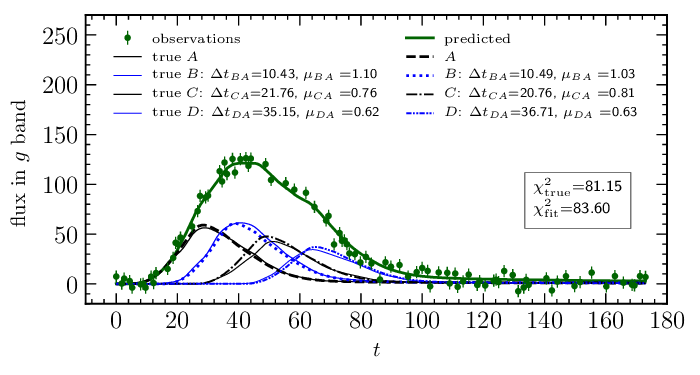}\\
	
	\includegraphics[width=0.47\textwidth]{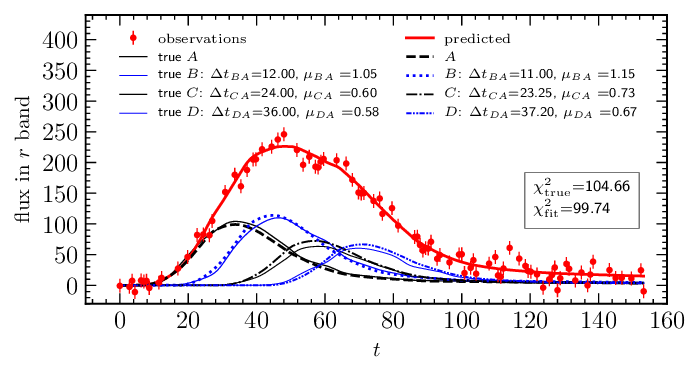}
	\includegraphics[width=0.47\textwidth]{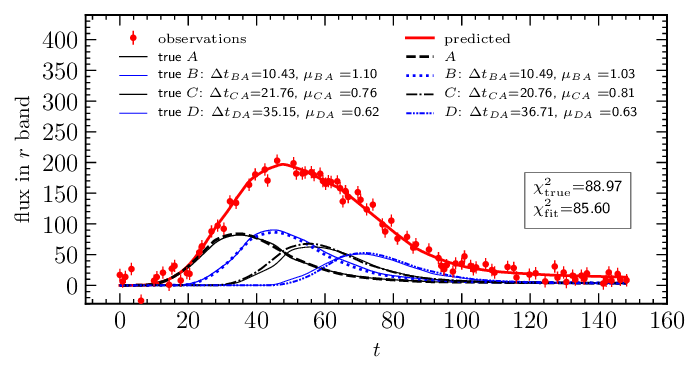}\\
	
	\includegraphics[width=0.47\textwidth]{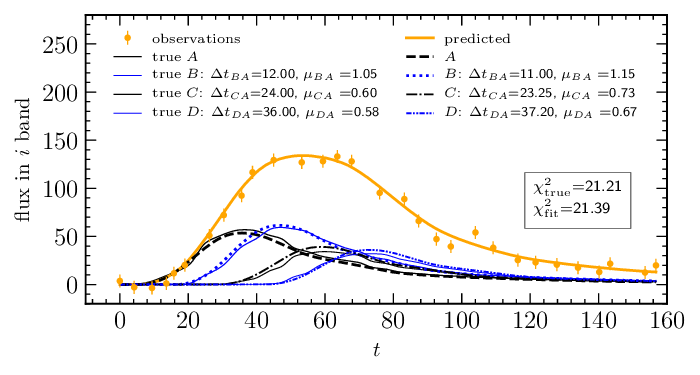}
	\includegraphics[width=0.47\textwidth]{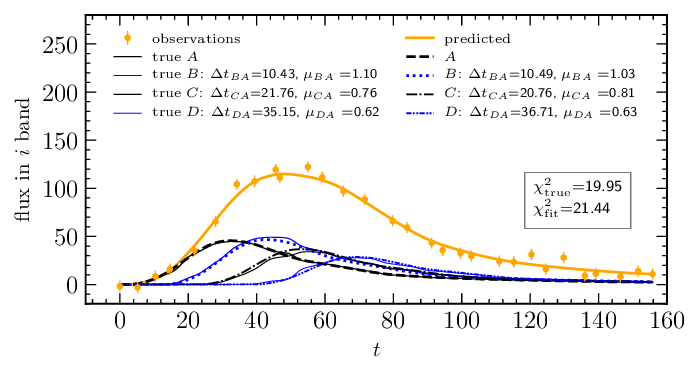}
	\caption{ 
		The simulated four-image data with $5\%$ noise level are shown for two systems, with equidistant $\Delta T=12$ time delays (left column) and randomly distributed time delays in the range $\dti=[10,14]$ (right column). The upper, middle, and bottom rows show the lightcurves in $g$, $r$, $i$ bands respectively. The 
		thin solid black and blue pairs of lines are the true light curves of individual images. The dashed, dotted, dash-dotted, and dash-double-dotted lines are the individual image flux reconstructions of our model using the best fit parameters. The extracted time delays and magnification ratios as well as their true counterparts are indicated in each panel. The $\chi^2$ values in the boxed texts indicate how well the reconstructed 
		lightcurves fit the observations in comparison to the true ones.
	} 
	\label{fig:l4imdT}
\end{figure*}

The right panel of Fig.~\ref{fig:sp4im} illustrates 
some time delay fits relative to the truth for these 
varying delay systems. The best fits are
scattered about the truth, with no strong correlation between different time delays in a system, i.e.\ if one is on the positive side, the others may be on the positive or negative side. (More quantitatively, the correlation coefficients are $\lesssim0.5$.)
As for the previous study of equal time delays, 
the fits are well consistent with the truth. 
Figure~\ref{fig:l4imdT} shows two examples of the four image 
fit to the observed lightcurve data, one with equal $\Delta T=12$ and 
one with time delays in the range $\dti=[10,14]$. 
The points with error bars are the 
observed data, with the thick solid line being the lightcurve 
constructed by the sum of the four image fits. The thin solid 
curves are the true image fluxes and the nonsolid curves the 
fit reconstructions -- remember, the only observable is the 
data points of the unresolved, summed lightcurve. The fit looks 
quite good, with excellent $\chi^2$ and accurate time delay 
fits. The magnification fits are less good but not unreasonable. 
Our main focus however is identifying that an unresolved 
system is in fact a lensed system -- out of one, many -- 
and how many images, rather than optimizing per se the time delay 
and magnification estimation.

%%%%%%%%%%%%%%%%%%%%%%%%%%%%%%%%%%%%%%%% 

%%%%%%%%%%%%%%%%%%%%%%%%%%% 
\section{How Many Images?} \label{sec:multi} 

While obtaining four good image fits for data simulated with four 
images (or two for two) is important and valuable, we will not 
know the true number of images 
behind observed data (unlike for resolved lensed 
systems). Therefore it is essential that we be able to distinguish 
systems with one (unlensed) vs two vs four images. Fitting a 
system with the incorrect number of images will likely lead to 
either degeneracies or biases. 

We have already explored this partially with 
Figure~\ref{fig:chi4vs2T}, and found robust distinction 
between four images and two images for $\Delta T\gtrsim8$ days. Now we examine this more thoroughly, forming a 
statistical confusion matrix between one (unlensed), 
two, and four image systems, for realizations of random 
time delays $\dti=[10,14]$ and magnifications. 

We simulate one, two, and four image systems, 100 of each, and 
fit every one with one, two, and four images (e.g.\ not just true two 
with fit two, but true two with fit one, two, and four).  
By examining the 
$\Delta\chi^2$ minima, and AIC, between the $N$ and $M$ image fits, we can decide 
whether we can robustly identify the number of images, and  
determine the time delays for the optimized case. The results of the 
fitting to simulations is summarized in the confusion matrix, 
with entries for each case where there are $N$ images simulated, 
i.e.\ truth, and $M$ images fit, for $N,M=1$, 2, 4. 
We assign the system to the highest number of images 
for which the AIC $<0$ and the fit time delays of the 
images exceed a given threshold, e.g.\ $\dtbf\ge10$ 
days. The fraction of true systems in each image 
category fit is given in the matrix entries. 

Table~\ref{tab:cf100} shows the results for our 
fiducial threshold, $\dtbf\ge10$ days. 
The confusion matrix is perfectly diagonal, showing 
purity of classification. As we have already seen 
in the previous sections, 
the time delay fit accuracy is excellent as well.

%%%%%%%%%%%%%%%%%%%%%%%%%%%%%%%%%%% 
\newlength{\colw}
\settowidth{\colw}{2 images }
\setlength\tabcolsep{0 pt}
\begin{table}
	\centering
	\renewcommand{\arraystretch}{1.2}
	\begin{tabular}{ll|M{\colw}M{\colw}M{\colw}|M{\colw}}    
		\multicolumn{2}{c}{}&   \multicolumn{4}{c}{\textbf{Predicted}}\\
		\multicolumn{2}{c}{}&\multicolumn{4}{c}{
			{\rotatebox[origin=c]{0}{~1 image}
			} {\rotatebox[origin=c]{0}{\ 2 images}
		} {\rotatebox[origin=c]{0}{\,4 images}
	}}\\
	\cline{3-5}
	\multirow{3}{*}{{\rotatebox[origin=c]{90}{\textbf{True}}
		}} & 
		1 image &\cellcolor{blue}\textcolor{white}1& 0 &0   \\ 
		%            \cline{3-5} 
		&   2 images\ \ &0 & \cellcolor{blue}\textcolor{white}1 &0   \\ 
		%        \cline{3-5} 
		&   4 images&0 & 0 &\cellcolor{blue}\textcolor{white}1   \\ \cline{3-5}
	\end{tabular}
	\caption{The confusion matrix showing the performance of our one, two, four image models on a simulation set consisting of 100 unlensed systems, 100 2-image and 100 4-image lensed systems. The lensed systems have true  adjacent time delays random in [10,14] and magnification ratios in [0.5,1.5]. 
		The threshold is $\dtbf\ge10$ days. 
	}
	\label{tab:cf100}
\end{table}

Studying the results as we reduce the threshold below 
our fiducial, we see in Table~\ref{tab:cf100dt8} that 
we still identify the number of images in the system 
perfectly for a threshold of $\dtbf\ge8$ days. Note we do have 10\% of two image systems 
that are fit as four image systems with {\it one\/} 
time delay greater than 8 days, but 0\% with 
{\it two\/} time delays greater than 8 days, and hence would 
not identify this as a four image system above 
threshold. It is not until the threshold is lowered 
to $\dtbf\ge6$ days that the confusion matrix becomes 
offdiagonal, as seen in Table~\ref{tab:cf100dt6}.

%%%%%%%%%%%%%%%%%%%%%%%%%%%%%%%%% 
\begin{table}
	\centering
	\renewcommand{\arraystretch}{1.2}
	\begin{tabular}{ll|M{\colw}M{\colw}M{\colw}|M{\colw}}    
		\multicolumn{2}{c}{}&   \multicolumn{4}{c}{\textbf{Predicted}}\\
		\multicolumn{2}{c}{}&\multicolumn{4}{c}{
			{\rotatebox[origin=c]{0}{~1 image}
			} {\rotatebox[origin=c]{0}{\ 2 images}
		} {\rotatebox[origin=c]{0}{\,4 images}
	}}\\
	\cline{3-5}
	\multirow{3}{*}{{\rotatebox[origin=c]{90}{\textbf{True}}
		}} & 
		1 image &\cellcolor{blue}\textcolor{white}1 & 0 &0   \\ 
		&   2 images\ \ &0 & \cellcolor{blue}\textcolor{white}1 &0   \\ 
		&   4 images&0 & 0 &\cellcolor{blue}\textcolor{white}1   \\ \cline{3-5}
	\end{tabular}
	\caption{As Table~\ref{tab:cf100} but with a 
		threshold of $\dtbf\ge8$ days.} 
	\label{tab:cf100dt8}
\end{table}

%%%%%%%%%%%%%%%%%%%%%%%%%%%%%%%%% 
\begin{table}
	\centering
	\renewcommand{\arraystretch}{1.2}
	\begin{tabular}{ll|M{\colw}M{\colw}M{\colw}|M{\colw}}    
		\multicolumn{2}{c}{}&   \multicolumn{4}{c}{\textbf{Predicted}}\\
		\multicolumn{2}{c}{}&\multicolumn{4}{c}{
			{\rotatebox[origin=c]{0}{~1 image}
			} {\rotatebox[origin=c]{0}{\ 2 images}
		} {\rotatebox[origin=c]{0}{\,4 images}
	}}\\
	\cline{3-5}
	\multirow{3}{*}{{\rotatebox[origin=c]{90}{\textbf{True}}
		}} & 
		1 image &\cellcolor{blue}\textcolor{white}{1} & 0 &0   \\ 
		&   2 images\ \ &0 & \cellcolor{NavyBlue}\textcolor{white}{0.97} &\cellcolor{LightCyan}0.03   \\ 
		&   4 images&0 & 0 &\cellcolor{blue}\textcolor{white}1   \\ \cline{3-5}
	\end{tabular}
	\caption{As Table~\ref{tab:cf100} but with a 
		threshold of $\dtbf\ge6$ days. 
	}
	\label{tab:cf100dt6}
\end{table}

Thus, the freeform method can successfully identify 
both the number of unresolved images, and the individual 
image time delays, for $\dtbf\gtrsim10$ days.

%%%%%%%%%%%%%%%%%%%%%%%%%%% 
\section{Conclusions} \label{sec:concl} 

Lensed Type Ia supernovae have the potential to become a 
significant new cosmological probe with the new generation 
of cosmic surveys. The time delays between images carry 
information on the scale and expansion history of the universe. 
However, not all multiply imaged SN will have {\it visibly\/} 
and {\it resolved\/} multiple images from which to measure the 
time delays. Images may be unresolved or blended, with a single 
measured lightcurve summing over the images. We investigate 
the problem of how to get out of one, many -- detecting the 
actual number of images and measuring their time delays. 

In Paper 1 we demonstrated one successful approach on two 
image systems. While it allowed wide variations in amplitude, 
it placed constraints on the shape of the lightcurve, correlated 
over its rise and fall. Here we include four image systems, 
where the variations in shape can be significant, so we 
develop and validate a new fitting method, essentially freeform 
in shape but more limited in amplitude variations. Neither 
method assumes a fixed template for the lightcurve. 

The freeform method performs quite well for time delays 
$\dt\gtrsim10$ days, and has reasonable behavior for 
$\dt\lesssim10$ days as well. We then address the question of 
multiplicity, distinguishing not only whether a source has 
been lensed, but accurately discerning the number of images, 
while simultaneously fitting for their time delays (and to 
a lesser extent their magnifications). The confusion matrix 
-- detailing whether a system is correctly identified or 
whether there are false negatives or positives -- is 
nearly diagonal, i.e.\ pure, for the freeform method for $\dt\gtrsim10$ 
days, and does well for shorter time delays as well. 

The goal of this research is to establish the ability 
to achieve out of one, many, to go as far as possible with 
unresolved but potentially blended lightcurves, in order to 
productively select systems for followup. That followup, 
e.g.\ obtaining resolved images through higher resolution 
imaging (and further ingredients such as for lens  
modeling) can turn the time delay systems into incisive 
cosmological probes. 

Our approach avoids fixed templates in the hope of 
eventually being applicable to a wide range of transients, 
beyond Type Ia supernovae. 
While the  crossing  technique in Paper 1 is particularly useful to deal with unclassified lensed transients, the 
freeform method used here is seen to be more precise and accurate for those suspected of being SN Ia that are of crucial importance in cosmology. A combination of these two approaches and crosschecks between their results offer 
the potential to deal with a broad range of transients. 
(While we focus on transients, combined lightcurves from 
quasars have been studied by, e.g., 
\citep{2011.04667,2101.11024,2101.11017,0501518,9601164}.) 

Future research could include recognition of lensing for diverse transients, fitting of partial lightcurves (e.g.\ early times) or different cadence (we studied the effect of S/N in Paper 1, and for this article chose a conservative noise level), etc. We will seek to improve further our approaches to classify various lensed transients and estimate their time delays.
An overall objective for the community is to determine accurate time delay distances for a large fraction of real systems (not just Type Ia supernovae).  While a large sample is desired, our focus here 
emphasized an unbiased pure sample over completeness. 
We present a model that is designed to be accurate for transients with {\it intrinsic\/} lightcurves within $\sim 0.1$ mag of a base shape at independent phases, and can characterize lensed systems when the unresolved {\it observed\/} summed lightcurves deviate more than 0.1 mag from that single unlensed source lightcurve.  
From simulations, we find that accepting SN with $\dtbf\gtrsim10$ effectively removes lensing false positives. 
We leave to further work the development of methods that identify more general systems with significant $\Delta t$, 
but a very different shape, i.e.\ deviating by more than 0.1 mag from the base shape. Comparison of fits of multiple models, such as the crossing method and the freeform method using base shapes from different source classes, may be used to identify systems needing further examination. 

In addition, Paper 3 (in draft) deals with the impact of microlensing 
on supernova time delay 
estimation; preliminary results show it appears tractable 
within the approaches of Paper 1 and this paper, while 
having two methods acts as a useful crosscheck.

%%%%%%%%%%%%
\section*{Acknowledgements}
We are grateful to Nazarbayev University Research 
Computing for providing computational resources 
for this work. This work was supported in part 
by the Energetic Cosmos Laboratory. 
AK and EL are supported in part by the U.S.\ Department of Energy, Office of Science, Office of High Energy Physics, under contract no.\ DE-AC02-05CH11231. 
AS would like to acknowledge the support by National Research Foundation of Korea NRF-2021M3F7A1082053, and the support of the Korea Institute for Advanced Study (KIAS) grant funded by the government of Korea. 

%%%%%%%%%%%%
\section*{Data Availability}
The data for the  300  simulated  lightcurves  of  1,  2,  and  4  image  systems, as used in the  confusion matrix results, are available on the GitHub repository \href{https://github.com/mdeatecl/LensedSN124imagesLCs}{https://github.com/mdeatecl/LensedSN124imagesLCs}. We include the \textsc{LCsimulator} code used to generate the lightcurves for this article in the same repository.


\bibliographystyle{mnras}
\bibliography{glsn}

\begin{thebibliography}{}
\makeatletter
\relax
\def\mn@urlcharsother{\let\do\@makeother \do\$\do\&\do\#\do\^\do\_\do\%\do\~}
\def\mn@doi{\begingroup\mn@urlcharsother \@ifnextchar [ {\mn@doi@}
  {\mn@doi@[]}}
\def\mn@doi@[#1]#2{\def\@tempa{#1}\ifx\@tempa\@empty \href
  {http://dx.doi.org/#2} {doi:#2}\else \href {http://dx.doi.org/#2} {#1}\fi
  \endgroup}
\def\mn@eprint#1#2{\mn@eprint@#1:#2::\@nil}
\def\mn@eprint@arXiv#1{\href {http://arxiv.org/abs/#1} {{\tt arXiv:#1}}}
\def\mn@eprint@dblp#1{\href {http://dblp.uni-trier.de/rec/bibtex/#1.xml}
  {dblp:#1}}
\def\mn@eprint@#1:#2:#3:#4\@nil{\def\@tempa {#1}\def\@tempb {#2}\def\@tempc
  {#3}\ifx \@tempc \@empty \let \@tempc \@tempb \let \@tempb \@tempa \fi \ifx
  \@tempb \@empty \def\@tempb {arXiv}\fi \@ifundefined
  {mn@eprint@\@tempb}{\@tempb:\@tempc}{\expandafter \expandafter \csname
  mn@eprint@\@tempb\endcsname \expandafter{\@tempc}}}

\bibitem[\protect\citeauthoryear{Bag, Kim, Linder  \& Shafieloo}{Bag
  et~al.}{2021}]{paper1}
Bag S.,  Kim A.~G.,  Linder E.~V.,   Shafieloo A.,  2021, \mn@doi [Astrophys.
  J.] {10.3847/1538-4357/abe238}, 910, 65

\bibitem[\protect\citeauthoryear{Barbary et~al.}{Barbary
  et~al.}{2020}]{sncosmo}
Barbary K.,  et~al., 2020, SNCosmo: Python library for supernova cosmology.
  Package version 2.1, \url {https://github.com/sncosmo/sncosmo}

\bibitem[\protect\citeauthoryear{Bellm et~al.,}{Bellm et~al.}{2018}]{ztf}
Bellm E.~C.,  et~al., 2018, \mn@doi [Pub. Ast. Soc. Pac.]
  {10.1088/1538-3873/aaecbe}, 131, 018002

\bibitem[\protect\citeauthoryear{Geiger \& Schneider}{Geiger \&
  Schneider}{1996}]{9601164}
Geiger B.,  Schneider P.,  1996, \mn@doi [Mon. Not. Roy. Astron. Soc.]
  {10.1093/mnras/282.2.530}, 282, 530

\bibitem[\protect\citeauthoryear{Goldstein, Nugent  \& Goobar}{Goldstein
  et~al.}{2019}]{goldstein}
Goldstein D.~A.,  Nugent P.~E.,   Goobar A.,  2019, \mn@doi [Astrophys. J.
  Suppl.] {10.3847/1538-4365/ab1fe0}, 243, 6

\bibitem[\protect\citeauthoryear{Guy et~al.}{Guy et~al.}{2007}]{salt}
Guy J.,  et~al., 2007, \mn@doi [Astron. Astrophys.]
  {10.1051/0004-6361:20066930}, 466, 11

\bibitem[\protect\citeauthoryear{Hazra \& Shafieloo}{Hazra \&
  Shafieloo}{2014}]{2014JCAP...01..043H}
Hazra D.~K.,  Shafieloo A.,  2014, \mn@doi [JCAP]
  {10.1088/1475-7516/2014/01/043}, 01, 043

\bibitem[\protect\citeauthoryear{Hsiao, Conley, Howell, Sullivan, Pritchet,
  Carlberg, Nugent  \& Phillips}{Hsiao et~al.}{2007}]{hsiao}
Hsiao E.~Y.,  Conley A.~J.,  Howell D.~A.,  Sullivan M.,  Pritchet C.~J.,
  Carlberg R.~G.,  Nugent P.~E.,   Phillips M.~M.,  2007, \mn@doi [Astrophys.
  J.] {10.1086/518232}, 663, 1187

\bibitem[\protect\citeauthoryear{{Huber} et~al.,}{{Huber}
  et~al.}{2022}]{2108.02789}
{Huber} S.,  et~al., 2022, \mn@doi [A\&A] {10.1051/0004-6361/202141956}, 658,
  A157

\bibitem[\protect\citeauthoryear{Liao, Shafieloo, Keeley  \& Linder}{Liao
  et~al.}{2020}]{liao}
Liao K.,  Shafieloo A.,  Keeley R.~E.,   Linder E.~V.,  2020, \mn@doi
  [Astrophys. J. Lett.] {10.3847/2041-8213/ab8dbb}, 895, L29

\bibitem[\protect\citeauthoryear{Linder}{Linder}{2004}]{linder04}
Linder E.~V.,  2004, \mn@doi [Phys. Rev. D] {10.1103/PhysRevD.70.043534}, 70,
  043534

\bibitem[\protect\citeauthoryear{Linder}{Linder}{2011}]{linder11}
Linder E.~V.,  2011, \mn@doi [Phys. Rev. D] {10.1103/PhysRevD.84.123529}, 84,
  123529

\bibitem[\protect\citeauthoryear{Millon et~al.}{Millon et~al.}{2020}]{millon}
Millon M.,  et~al., 2020, \mn@doi [Astron. Astrophys.]
  {10.1051/0004-6361/201937351}, 639, A101

\bibitem[\protect\citeauthoryear{Pierel, Rodney, Vernardos, Oguri, Kessler  \&
  Anguita}{Pierel et~al.}{2021}]{2010.12399}
Pierel J. D.~R.,  Rodney S.,  Vernardos G.,  Oguri M.,  Kessler R.,   Anguita
  T.,  2021, \mn@doi [Astrophys. J.] {10.3847/1538-4357/abd8d3}, 908, 190

\bibitem[\protect\citeauthoryear{Pindor}{Pindor}{2005}]{0501518}
Pindor B.,  2005, \mn@doi [Astrophys. J.] {10.1086/430048}, 626, 649

\bibitem[\protect\citeauthoryear{{Refsdal}}{{Refsdal}}{1964}]{refsdal}
{Refsdal} S.,  1964, \mn@doi [Mon. Not. Roy. Astron. Soc.]
  {10.1093/mnras/128.4.295}, 128, 295

\bibitem[\protect\citeauthoryear{Rodney, Brammer, Pierel, Richard, Toft,
  O'Connor, Akhshik  \& Whitaker}{Rodney et~al.}{2021}]{2106.08935}
Rodney S.~A.,  Brammer G.~B.,  Pierel J. D.~R.,  Richard J.,  Toft S.,
  O'Connor K.~F.,  Akhshik M.,   Whitaker K.,  2021, \mn@doi [Nature Astronomy]
  {10.1038/s41550-021-01450-9}, 5, 1118

\bibitem[\protect\citeauthoryear{Shafieloo}{Shafieloo}{2012a}]{2012JCAP...05..024S}
Shafieloo A.,  2012a, \mn@doi [JCAP] {10.1088/1475-7516/2012/05/024}, 05, 024

\bibitem[\protect\citeauthoryear{Shafieloo}{Shafieloo}{2012b}]{2012JCAP...08..002S}
Shafieloo A.,  2012b, \mn@doi [JCAP] {10.1088/1475-7516/2012/08/002}, 08, 002

\bibitem[\protect\citeauthoryear{Shafieloo, Clifton  \& Ferreira}{Shafieloo
  et~al.}{2011}]{2011JCAP...08..017S}
Shafieloo A.,  Clifton T.,   Ferreira P.~G.,  2011, \mn@doi [JCAP]
  {10.1088/1475-7516/2011/08/017}, 08, 017

\bibitem[\protect\citeauthoryear{Shajib et~al.}{Shajib et~al.}{2020}]{shajib}
Shajib A.~J.,  et~al., 2020, \mn@doi [Mon. Not. Roy. Astron. Soc.]
  {10.1093/mnras/staa828}, 494, 6072

\bibitem[\protect\citeauthoryear{Shu, Belokurov  \& Evans}{Shu
  et~al.}{2021}]{2011.04667}
Shu Y.,  Belokurov V.,   Evans N.~W.,  2021, \mn@doi [Mon. Not. Roy. Astron.
  Soc.] {10.1093/mnras/stab241}, 502, 2912

\bibitem[\protect\citeauthoryear{Springer \& Ofek}{Springer \&
  Ofek}{2021a}]{2101.11017}
Springer O.~M.,  Ofek E.~O.,  2021a, \mn@doi [Mon. Not. Roy. Astron. Soc.]
  {10.1093/mnras/stab1600}, 506, 864

\bibitem[\protect\citeauthoryear{Springer \& Ofek}{Springer \&
  Ofek}{2021b}]{2101.11024}
Springer O.~M.,  Ofek E.~O.,  2021b, \mn@doi [Mon. Not. Roy. Astron. Soc.]
  {10.1093/mnras/stab2432}, 508, 3166

\bibitem[\protect\citeauthoryear{{Stan Development Team}}{{Stan Development
  Team}}{2021}]{stan}
{Stan Development Team} 2021, CMDStan: the shell interface to Stan. Package
  version 2.27, \url {http://mc-stan.org/}

\bibitem[\protect\citeauthoryear{Treu \& Marshall}{Treu \&
  Marshall}{2016}]{treumar}
Treu T.,  Marshall P.~J.,  2016, \mn@doi [Astron. Astrophys. Rev.]
  {10.1007/s00159-016-0096-8}, 24, 11

\bibitem[\protect\citeauthoryear{Wong et~al.}{Wong et~al.}{2020}]{wong}
Wong K.~C.,  et~al., 2020, \mn@doi [Mon. Not. Roy. Astron. Soc.]
  {10.1093/mnras/stz3094}, 498, 1420

\makeatother
\end{thebibliography}

%%%%%%%%%%%%%%%%%%%%%%%%%%%%%%%%%%%%%%%%%%%%%%%%%%

%%%%%%%%%%%%%%%%% APPENDICES %%%%%%%%%%%%%%%%%%%%%

%%%%%%%%%%%%
\begin{appendix}

%%%%%%%%%%%%%%%%%%%%%%%% 
\section{Testing Freeform with SALT2} \label{sec:apxsalt} 

The Hsiao template is used in two ways: to generate 
the simulated systems, and also as the base from which 
the freeform method starts in allowing shape deviations. 
We want to assess whether this affects the fits, despite 
the addition of 50-70 free parameters allowing a free form. 
To do so, we generate systems using the SALT2 lightcurve 
template \cite{salt} instead, and fit them with the freeform 
method acting on the Hsiao base, 
i.e.\ Eq.~\eqref{eq:unlens}.

%%%%%%%%%%%%%%%%%%%%%%%%%%%%%%%%%%%%% 
\begin{figure*}
	\centering
	\includegraphics[width=0.45\textwidth]{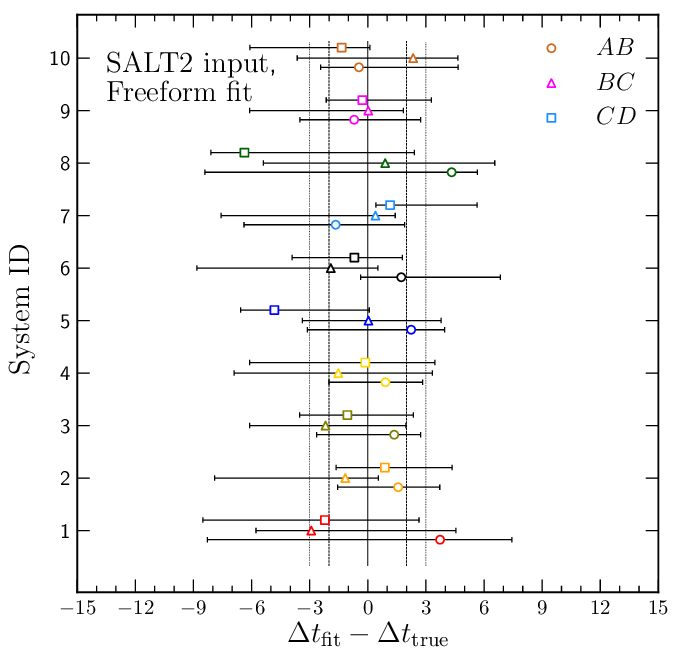}
	\includegraphics[width=0.45\textwidth]{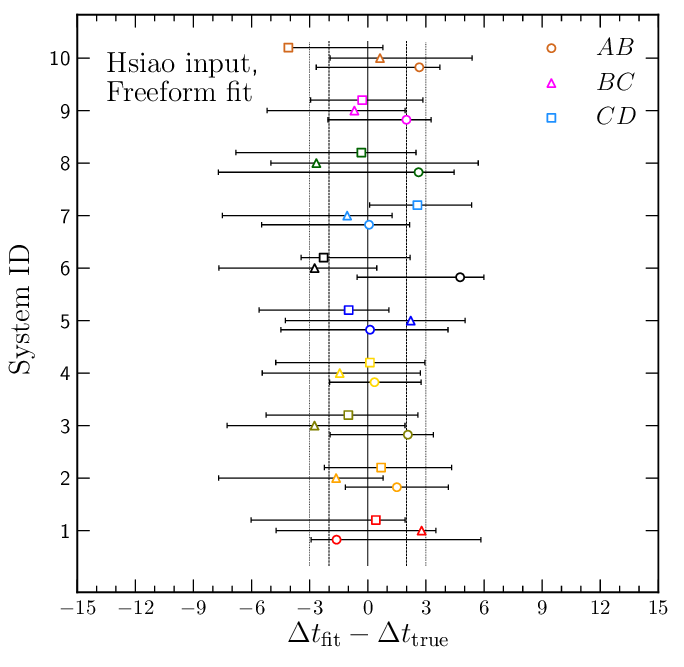}
	\caption{The time delay deviations from the truth in four-image systems generated using the SALT2 (left panel) or Hsiao (right panel; as the right panel in Fig.~\ref{fig:sp4im})  templates and fitted using the freeform method with the Hsiao base. The time delays are randomly chosen from the $\dti=[10,14]$ interval. 
		The freeform fit works even with 
		a different supernova input form. Error bars show the  
		95\% confidence intervals.
	} 
	\label{fig:salthsiao}
\end{figure*}

Figure~\ref{fig:salthsiao} shows that the freeform fit 
works comparably well whether the data is generated 
based on SALT2 or Hsiao. 
As a further check we 
have verified that the clear distinction between 
four image and two image systems continues to hold 
with the freeform method even when generated with SALT2.

%%%%%%%%%%%%%%%%%%%%%%%%%%%%%%% 
\section{Estimating 2-image Time Delays Below 8 days} \label{sec:aless10} 

\begin{figure}
	\centering
	\includegraphics[width=\columnwidth]{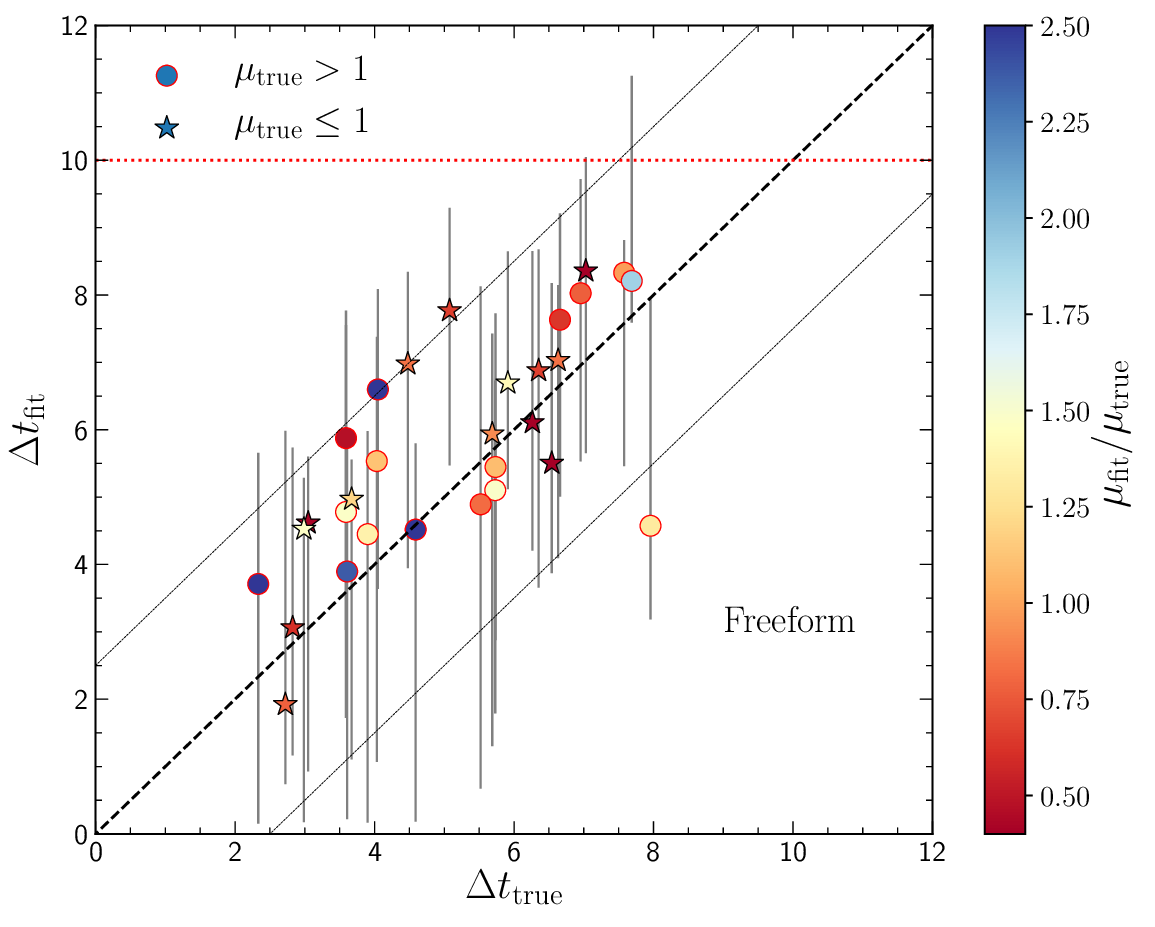} 
	\caption{Time delay fits are given for 30 two image systems, 
		showing that $\dtbf\approx\dtt$ even in the regime 
		$2\le \dtt<8$ days, below the threshold used in the main text. 
		The dashed diagonal line gives $\dtbf=\dtt$, 
		while the dotted diagonals give $\pm2.5$ days around this. Error bars show the  
		95\% confidence intervals.
	}
	\label{fig:less10}
\end{figure}

As seen in Fig.~\ref{fig:ablindtest} for two image systems, 
the freeform method continues to yield $\dtbf\approx\dtt$ 
below the fiducial threshold of $\dtbf=10$ days. Although 
we do not use systems with shorter time delays in the main 
text, here we explore two image time delays with 
$2\le \dtt<8$ days. Figure~\ref{fig:less10} shows that 
the fits continue to follow $\dtbf\approx\dtt$, with 
scatter of approximately two days (comparable to the 
observation cadence). However, also as seen in 
Fig.~\ref{fig:ablindtest}, we expect unlensed systems 
to ``upscatter'' into this region of $\dtbf$ and so 
to avoid false positives we conservatively do not use a 
lower threshold  $\dtbf<10$.

%%%%%%%%%%%%%%%%%%%%%%%%%%%%%% 
\end{appendix}

%%%%%%%%%%%%%%%%%%%%%%%%%%%%%%%%%%%%%%%%%%%%%%%%%%

% Don't change these lines
\bsp	% typesetting comment
\label{lastpage}
\end{document}